\documentclass[lettersize,journal]{IEEEtran}
\usepackage{amsmath,amsfonts}
\usepackage{algorithmic}
\usepackage{array}
\usepackage[caption=false,font=normalsize,labelfont=sf,textfont=sf]{subfig}
\usepackage{textcomp}
\usepackage{cite}
\usepackage{stfloats}
\usepackage{url}
\usepackage{verbatim}
\usepackage{braket}
\usepackage{graphicx}
\graphicspath{ {./figures/} }
\usepackage[T1]{fontenc}
\usepackage{pdfpages}
\usepackage[utf8]{inputenc}
\usepackage{authblk}
\def\BibTeX{{\rm B\kern-.05em{\sc i\kern-.025em b}\kern-.08em
    T\kern-.1667em\lower.7ex\hbox{E}\kern-.125emX}}
\usepackage{balance}
\begin{document}
\title{Q-PnV: A Quantum Consensus Mechanism for Security Consortium Blockchains}

\author[1]{Jianming Lin}
\author[1,2,*]{Hui Li}
\author[1]{Hongjian Xing}
\author[3]{Runhuai Huang}
\author[4]{Weixiang Huang}
\author[5]{Shaowen Deng}
\author[4]{Yanping Zhang}
\author[3]{Weimin Zeng}
\author[5]{Ping Lu}
\author[6]{Xiyu Wang}
\author[7,2]{Tao Sun}
\author[8]{Xiongyan Tang}

\affil[1]{School of Electronic and Computer Engineering, Peking University Shenzhen Graduate School, Shenzhen 518055, China.}
\affil[2]{School of Computer Science, Fuyao University of Science and Technology, Fuzhou 350109, China.}
\affil[3]{China Telecom Cloud Technology Co., Ltd., Beijing 100007, China.}
\affil[4]{Capability \& Platform Business Dept., China Mobile Internet Co., Ltd., Guangzhou 510640, China.}
\affil[5]{Cloud-Network of China Unicom Global Limited, Hongkong 510031, China.}
\affil[6]{Zhongxing Telecom Equipment (ZTE) Corporation, Shenzhen 518057, China.}
\affil[7]{Management Service Center of Shenzhen University Town, Shenzhen 518055, China.}
\affil[8]{China Unicom Research Institute, Beijing 102600, China.}
\affil[*]{Correspondence: lih64@pkusz.edu.cn; 202401310009@fyust.org.cn}

\markboth{}%
{How to Use the IEEEtran \LaTeX \ Templates}

\maketitle

\begin{abstract}
Due to the rapid development of quantum computing, many classical blockchain technologies are now considered insecure. The emergence of quantum blockchain holds promise for addressing this issue. Various quantum consensus algorithms have been proposed so far, but there has not yet been a quantum consensus algorithm tailored specifically for consortium blockchain scenarios. In this paper, we propose a novel quantum consensus mechanism, named Q-PnV. This consensus mechanism is based on the classical Proof of Vote (PoV), integrating quantum voting, quantum digital signature and quantum random number generators (QRNGs). By combining Q-PnV with a quantum blockchain using weighted hypergraph states, we propose a comprehensive quantum blockchain solution for consortium blockchain scenarios. Compared to the classical method, the quantum blockchain based on Q-PnV can resist quantum attacks and shows significant improvements in security and fairness, making it better suit-ed for the future quantum era.
\end{abstract}


\section{Introduction}
\IEEEPARstart{B}{lockchain} is a decentralized distributed ledger technology that is widely used in various fields, including finance, supply chain management, healthcare, and the Internet of Things. With the rapid development of quantum computing, the security of classical blockchain becomes severely threatened. The emergence of Shor's algorithm \cite{shor_algorithms_1994} and Grover's algorithm \cite{grover_quantum_1997} has raised concerns about the potential destruction of the underlying encryption systems of blockchain. Digital signatures and hash functions in classical blockchain appear vulnerable in the face of quantum computing. Therefore, it is particularly important to develop blockchain technologies that can resist quantum attacks.

To address the quantum challenges in blockchain, two different approaches have been proposed \cite{yang_survey_2024}. The first approach is post-quantum blockchain, which uses post-quantum cryptographic algorithms to replace classical cryptographic algorithms in the blockchain system. This method is compatible with existing cryptographic infrastructures and can reach high secret-key rates over relatively long distances. However, its security may be compromised by currently unknown algorithms in the future, as it can only resist known quantum attacks \cite{cao_evolution_2022}. The second approach is quantum blockchain, which utilizes the principles of quantum mechanism to establish a completely new blockchain system that is considered absolutely secure. Quantum blockchain can be a hybrid architecture that combines classical and quantum components or an entirely quantum-based system \cite{yang_survey_2024}. A hybrid quantum blockchain mainly refers to a blockchain system based on quantum key distribution (QKD). In 2018, Kiktenko et al. proposed a quantum-secured blockchain that replaces digital signatures in classical blockchain with QKD \cite{kiktenko_quantum-secured_2018}, while blocks are generated through other decentralized methods. A fully quantum blockchain, on the other hand, leverages the powerful capabilities of quantum computers and quantum networks to the fundamental structure of blockchain. In 2019, Rajan and Visser proposed a quantum blockchain scheme using temporal entangled GHZ states \cite{rajan_quantum_2019}, pioneering this field. However, their scheme remains theoretical without any implementation methods or designs for real quantum devices or simulators. In 2020, Banerjee et al. proposed another quantum blockchain scheme based on weighted hypergraph states \cite{banerjee_quantum_2020}, which, despite its lack of completeness, presented specific implementation methods. Subsequent research has refined this scheme in various aspects \cite{li_efficient_2022, orts_improving_2023}. In 2022, Nilesh and Panigraphi \cite{nilesh_quantum_2022} proposed a fully quantum blockchain based on a generalized Gram-Schmidt procedure utilizing dimensional lifting, which is currently the most complete quantum blockchain scheme, but its completeness means it cannot yet be implemented or simulated on any quantum device.

Apart from the construction of quantum blockchains, some research focus on quantum consensus algorithms, applying quantum technology to consensus algorithms, such as quantum zero-knowledge proof \cite{wen_blockchain_2022}, quantum teleportation \cite{wen_blockchain_2022-1}, quantum random numbers \cite{wang_consensus_2022}, and quantum voting \cite{li_efficient_2022}. These consensus algorithms can be combined with classical or quantum blockchain to form a complete quantum blockchain solution. Some of these consensus algorithms are entirely new, based on quantum technology, while others modify and improve classical consensus algorithms using quantum properties. 

Blockchain can generally be divided into three types: public blockchain, private blockchain, and consortium blockchain. In public blockchains, due to the large number of nodes and the lack of trust among participants, the system requires extremely high stability and fault tolerance, while also striving to achieve decentralization as much as possible. This imposes significant demands on the underlying network hardware. However, current quantum Internet remains in its early stages. Quantum computers are costly and lack stability, making them challenging to deploy effectively in large-scale distributed networks. In contrast, the characteristics of consortium blockchains make them more suitable for the application of quantum technology at this stage. The nodes in a consortium blockchain are usually composed of known and trusted entities, and the number of nodes is limited. This makes the deployment, management, and configuration of hardware resources within the network relatively simple. Additionally, the participants in a consortium blockchain typically possess significant economic resources, allowing them to afford the high costs associated with quantum hardware. These factors make it feasible to implement quantum consensus mechanisms within consortium blockchains. Therefore, applying quantum technology to the consensus mechanism in consortium blockchains represents a viable pathway for the gradual development of quantum blockchains.

Proof of Vote (PoV) \cite{li_proof_2017,li_pov_2020} is a consensus mechanism specifically designed for consortium blockchains, known for its low latency and high performance in classical computing environments. PnV \cite{bai_parallel_2021,wang_pnv_2024}, an enhanced version of PoV, is a parallel fusion protocol that combines proof-based and voting-based consensus features, significantly improving throughput while maintaining acceptable latency. It excels in efficiency and scalability. However, like other classical consensus mechanisms, both PoV and PnV are vulnerable to quantum threats. In this paper, we propose a quantum-enhanced version of PoV and PnV, transforming them into a quantum consensus mechanism specifically designed for consortium blockchains. By integrating quantum technologies, we aim to strengthen security and resilience against quantum attacks, while retaining the performance benefits of the classical approaches. Considering that PnV is a better version, we name this quantum consensus mechanism Q-PnV.

The main contributions of this paper are summarized as follows:
\begin{itemize}
	\item[$\bullet$] We modify and enhance the classical proof-based and voting-based consensus mechanisms using quantum technology, introducing a quantum consensus mechanism named Q-PnV for consortium blockchains. This mechanism significantly improves security and fairness, particularly in the context of quantum threats. 
	\item[$\bullet$] We integrate Q-PnV with a quantum blockchain frame-work based on weighted hypergraph states, presenting a comprehensive quantum blockchain solution tailored for consortium blockchain scenarios. This solution enhances security and fairness, making it more suitable for addressing quantum threats in consortium blockchain environments.
\end{itemize}

The remainder of this paper is organized as follows: Section \uppercase\expandafter{\romannumeral2} provides an overview of classical blockchain and its vulnerabilities to quantum threats, along with a brief explanation of the quantum technologies and principles relevant to this work. Section \uppercase\expandafter{\romannumeral3} introduces the Q-PnV consensus mechanism, developed by enhancing the classical consensus using quantum technology. Section \uppercase\expandafter{\romannumeral4} presents the integration of Q-PnV with a quantum blockchain based on weighted hypergraph states, offering a complete quantum consortium blockchain solution. Section \uppercase\expandafter{\romannumeral5} evaluates the security and fairness of the proposed scheme.

\section{Preliminary}
In this section, we provide an introduction to the foundational concepts and related work on both classical blockchain and quantum blockchain technologies.

\subsection{Classical Blockchain and Consensus}
1) Structure of Classical Blockchain: Blockchain is essentially a decentralized, distributed database that uses consensus mechanisms and classical cryptographic techniques to ensure the security and immutability of data within the network. Transactions are the most common type of data in a blockchain, which is why blockchain is also considered a distributed ledger. All transactions that occur in the network are packaged into blocks, and these blocks are combined to form a chain-like structure. Each block includes multiple transactions along with a block header that contains the hash of the previous block, as illustrated in Fig.1.

\begin{figure}[b]
	\centering
	\includegraphics[width=0.48\textwidth]{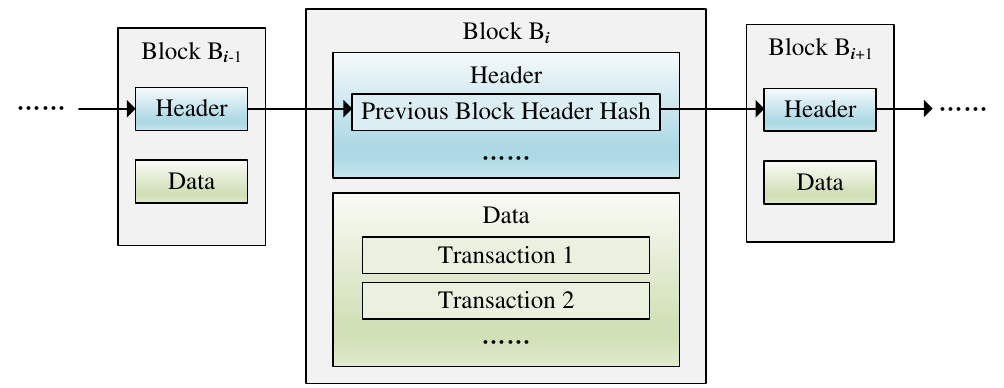}
	\caption{The simplified blockchain data structure.} 
	\label{Fig1}
\end{figure}

\begin{figure*}[htb]
	\centering
	\includegraphics[width=0.7\textwidth]{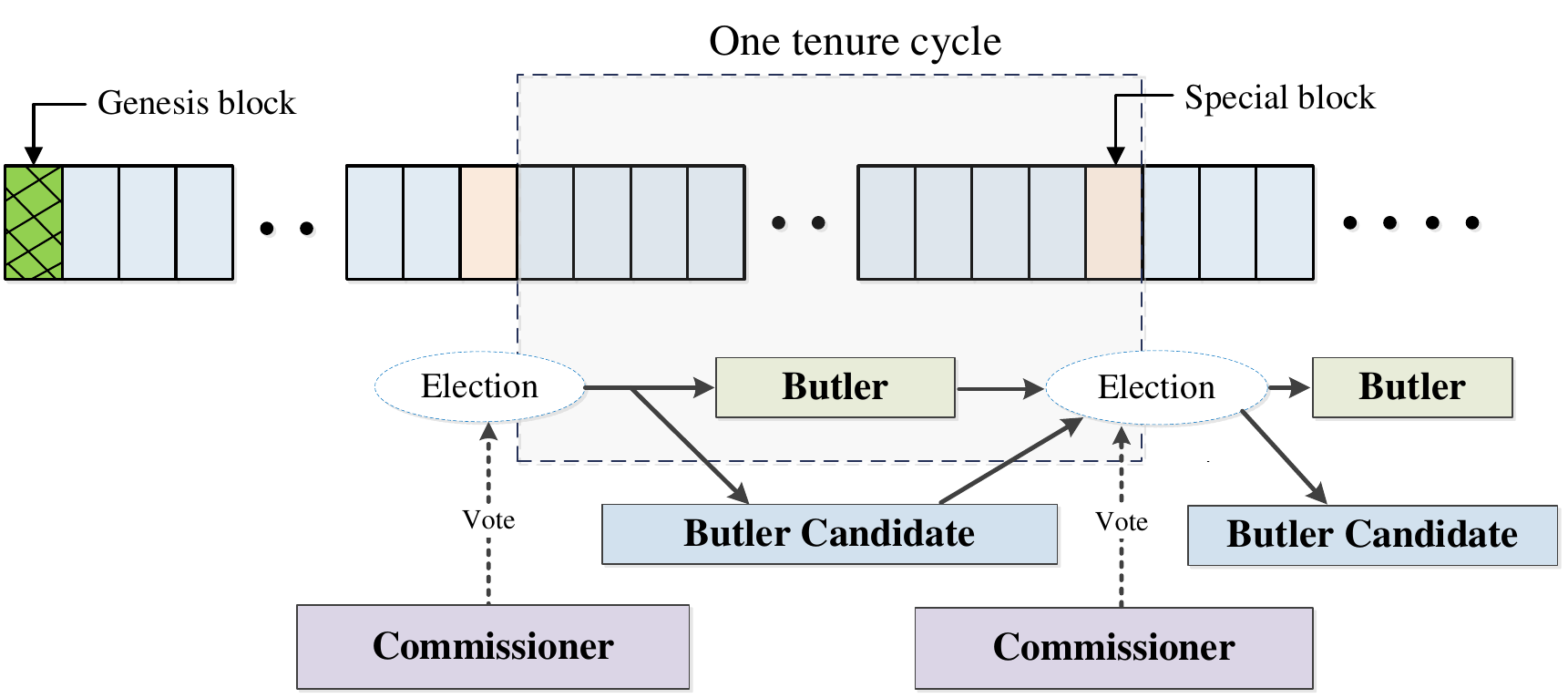}
	\caption{One tenure cycle process in PoV \cite{li_proof_2017}.} 
	\label{Fig2}
\end{figure*}

2) Classical Consensus based on Proof and Voting: Blockchain consensus algorithms can be categorized from various perspectives \cite{li_CAP_2024}. One classification divides them into voting-based, proof-based, and hybrid consensus mechanisms. They can also be distinguished by fault tolerance as Byzantine Fault Tolerant (BFT) and non-BFT consensus algorithms. Based on deployment, consensus mechanisms are classified into public blockchain, consortium blockchain, and private blockchain consensus. In terms of consistency, they can be categorized into strongly consistent, weakly consistent, or final consistent consensus. Additionally, based on the network model assumptions, consensus algorithms fall into three types: synchronous, partially synchronous, and asynchronous network consensus, with the partially synchronous model being the most commonly used in blockchain consensus mechanisms.
 
PoV is the first consensus mechanism to integrate proof and voting \cite{li_proof_2017,li_pov_2020}, which eliminates the need for computing power, leading to reduced latency and increased throughput. From a security perspective, PoV requires just over half of the voters to function correctly and at least one honest bookkeeper to ensure the seamless operation of the consensus process without forks. The network communication complexity is $O(N_v)$, which is significantly lower than PBFT’s $O(N_{all}^2)$ \cite{castro_practical_1999, sukhwani_performance_2017}. In 2019, PnV \cite{bai_parallel_2021, wang_pnv_2024} was proposed as an improved version of PoV, enhancing both the data structure and the consensus process by enabling multiple bookkeepers to generate blocks simultaneously. This optimization significantly boosts the overall throughput of the blockchain system. Subsequently, Lightweight PnV \cite{wang_data_2021} and Mimic PnV \cite{xiao_optimizing_2022} consensus mechanism were successively proposed.

In the design of PoV, nodes in the network are divided into four roles: commissioner, butler, butler candidate and ordinary user. Commissioners are members of the consortium committee, with the rights to recommend, vote and evaluate butlers. Commissioners are also responsible for verifying and forwarding blocks and transactions. Transactions in valid blocks must receive more than $50\%$ approval votes from commissioners. Butlers are bookkeeper nodes in the consortium blockchain, granted the authority to generate blocks. Nodes wishing to become butlers must first become butler candidates. Ordinary users and commissioners can apply to become butler candidates. A node can simultaneously hold the dual roles of commissioner and butler.

The block generation cycle in PoV is divided into rotation cycle and tenure cycle. A rotating butler is randomly selected in each rotation cycle, who is responsible for generating a block. Multiple rotation cycles constitute a tenure cycle. The last round of each tenure cycle is used to elect butlers. And the last block of each tenure cycle is used to record election results and related information, which is called a special block. Fig.2 illustrates the process within one tenure cycle.

3) Quantum Threats in PoV: With the rapid development of quantum technology, the security of classical blockchain is severely threatened. PoV stipulates that transactions such as voting and recommendations in the consensus process must be signed. The digital signatures used in classical blockchain are based on certain mathematical problems, such as large integer factorization. However, Shor's quantum algorithm can solve these problems in polynomial time \cite{shor_algorithms_1994}, rendering classical digital signature insecure. Attackers can easily obtain the private keys of commissioners using Shor's algorithm, thereby disrupting the voting and election processes. Additionally, the rotating butler in PoV is selected by hashing the signatures and timestamps. Grover's quantum search algorithm can accelerate the inverse computation of hash functions \cite{grover_quantum_1997}, reducing its time complexity to $O(\sqrt{n})$. Attackers can use Grover's algorithm to consistently maintain the block-producing rights, severely undermined the decentralization of the blockchain.

\subsection{Quantum Blockchain}

\begin{figure}[b]
	\centering
	\includegraphics[width=0.5\textwidth]{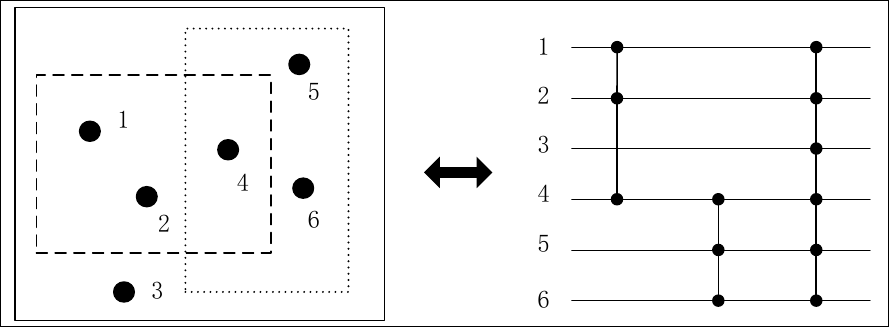}
	\caption{A quantum hypergraph state with seven vertices and three hypergraphs.} 
	\label{Fig3}
\end{figure}

1) Quantum Weighted Hypergraph: Mathematically, a hypergraph refers to a special version of a graph. In an ordinary graph, an edge connects two nodes, whereas in a hypergraph, an edge can connect two or more nodes. Such an edge is called a hyperedge. Quantum hypergraph applies the definition of hypergraph to multipartite entangled qubits \cite{rossi_quantum_2013}, where the nodes correspond to qubits and the hyperedges represent the connections between these qubits. An example of a hypergraph and its quantum equivalent is shown in Fig.3.

The qubits corresponding to each hyperedge are entangled by using Controlled-Z ($C^kZ$) quantum gates, where $k$ represents the number of control bits in the quantum gate. The initial state of each qubit is $\ket{+}=\ket{0}+\ket{1}$. The resulting quantum state of the circuit shown in Fig.3 will then be:
\begin{align}
	\ket{\psi}=C_{(1,2,4)}^2Z C_{(4,5,6)}^2Z C_{(1,2,3,4,5,6)}^5Z \ket{+}
\end{align}

The concept of a hypergraph can be extended to a weighted hypergraph, where each edge is assigned a weight \cite{rossi_quantum_2013}. The definition of an N-qubit weighted hypergraph state is \cite{qu_encoding_2013}:
\begin{align}
	\ket{\psi} = \frac{1}{\sqrt{2^N}} \sum_{x\in(0,1)^N} e^{i\pi f(x)} \ket{x}
\end{align}
where $\ket{x}$ refers to the computational basis and $f(x)$ represents the corresponding weight value, $f(x)\in(0,1)$.

2) Quantum Blockchain Using Weighted Hypergraph States: In 2020, Banerjee et al. proposed a quantum blockchain scheme \cite{banerjee_quantum_2020}, in which classical information is encoded into the phase of a single qubit, with each block using only one qubit. These qubits serve as the vertices of the corresponding hypergraph state, where they are entangled to connect and form a chain structure. The specific steps of the scheme are as follows.

A quantum block encodes classical information p into the phase of a qubit, with one qubit corresponding to one block. The peer who made the block initializes the qubit as $\ket{+}$ and introduces $p$ as:
\begin{align}
	\ket{\psi_1} = S(p)\ket{+} = \frac{1}{\sqrt{2}} \begin{pmatrix}
		1 & 0 \\
		0 & e^{i\theta_p}
	\end{pmatrix} \ket{+} =  \frac{\ket{0}+e^{i\theta_p}\ket{1}}{\sqrt{2}}
\end{align}

It is required in the protocol that
\begin{align}
	\theta_p\in(0,\frac{\pi}{2})
\end{align}
and
\begin{align}
	\sum_{i}\theta_{p_i}<\frac{\pi}{2}
\end{align}
where $i$ refers to the number of blocks. Then the phase angles of the quantum blocks are set to be equiproportional:
\begin{align}
	\theta_{p_i} = \frac{1}{2(i-1)} \theta_{p_1}
\end{align}

Considering that series $\sum_{i=1}^\infty \theta_{p_i}$  converges to $2\theta_{p_1}$, the value of the first block $\theta_{p_1}$ should less than $\pi/4$. Eq.6 shows that once the first quantum block is determined, subsequent blocks can be determined by the proportional relationship between their phase angles. Finally, all the quantum blocks are entangled into a weighted hypergraph state to form a quantum blockchain through a quantum circuit shown in Fig.4.

\begin{figure}[t]
	\centering
	\includegraphics[width=0.5\textwidth]{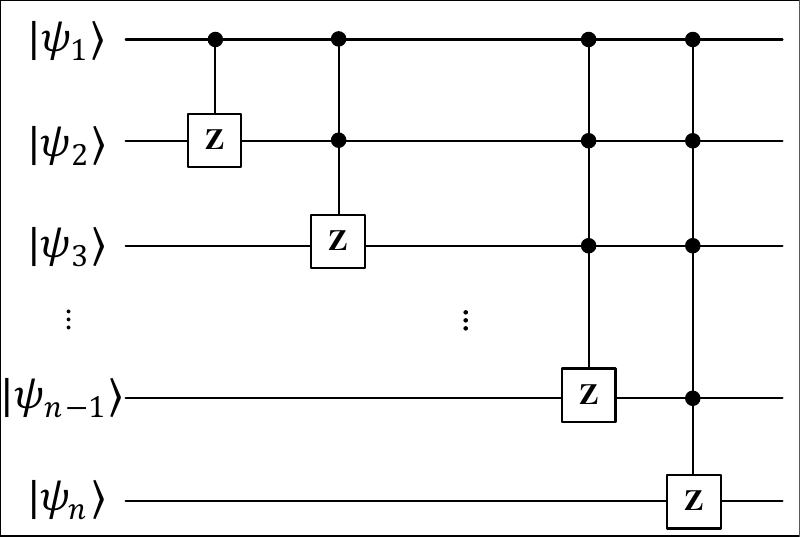}
	\caption{Circuit diagram of a quantum blockchain using weighted hypergraph states with N blocks.} 
	\label{Fig4}
\end{figure}

\section{Quantum Consensus Q-PnV}
PoV and its enhanced versions are specifically designed for consortium blockchain, and the same applies to Q-PnV. This consensus mechanism improve PoV by using quantum technology which ensures security and fairness even in the quantum era.

\subsection{Enabling Technologies of Q-PnV}
Quantum Internet: Classical blockchain relies on the classical Internet, and quantum blockchain similarly requires the use of the quantum Internet. Quantum Internet is a new type of network that relies on quantum technology and classical technology for information transmission and processing \cite{singh_quantum_2021}. In 2018, Wehner et al. published a forward-looking paper on the roadmap of quantum Internet \cite{wehner_quantum_2018}, dividing the development of quantum Internet into six stages. According to the content of the paper and the current state of development, the quantum Internet is still in its infancy. Core technologies of the quantum Internet, such as quantum memory and quantum repeater, are still in the open exploration stage of theoretical and experimental research. While some performance indicators are continuously improving, there is still a significant gap from practical levels. 

Although the current development of the quantum Internet does not support large-scale deployment of quantum blockchain, research on quantum blockchain is very necessary to avoid attacks on classical blockchain by quantum computing. In this paper, we assume that a distributed interconnected quantum network has already been realized, and operations such as the preparation and distribution of qubits in the network meet practical levels.

2) Two Types of Quantum Multiparticle Entangled States in Voting Process: Voting is widely used in many consensus mechanism. Classical voting protocols are usually based on public-private key cryptographic algorithms, which makes them face significant security threats due to quantum computing. Quantum voting is a new method that utilizes quantum technologies to achieve secure, efficient, and fair voting. In 2016, Wang et al. proposed a quantum voting protocol \cite{wang_self-tallying_2016} firstly possesses privacy, self-tallying, non-reusability, verifiability and fairness at the same time by using two special types of multipartite entangled states $\ket{X_n}$ and $\ket{S_n}$. The protocol exploits their different properties in computational basis $\{\ket{j}_C, j=0,1,\cdots,m-1\}$ and Fourier basis $\{\ket{j'}_F, j'=0,1,\cdots,m-1\}$ to achieve security verification. The $m$-level $n$-particle state $\ket{X_n}$ in computational basis is defined as:
\begin{align}
	\ket{X_n}_c = \frac{1}{\sqrt{m^{n-1}}} \sum_{\sum_{i=0}^{n-1} j_i\mod m=0} \ket{j_0}_c \ket{j_1}_c \cdots \ket{j_{n-1}}_c
\end{align}

And after the quantum Fourier transform, it can be rewritten as:
\begin{align}
	\ket{X_n}_F = \frac{1}{\sqrt{m}} \sum_{j'=0}^{m-1} \ket{j'}_{0,F} \ket{j'}_{1,F} \cdots \ket{j'}_{n-1,F}
\end{align}
where
\begin{align}
	\ket{j'}_F = \mathcal{F} \ket{j}_C = \frac{1}{\sqrt{m}} \sum_{k=0}^{m-1} e^{\frac{2\pi i}{m} jk} \ket{k}_C
\end{align}

According to the state of $\ket{X_n}$, it has the following properties: when measured in the computational basis, the sum over all the $n$ measurement outcomes modulo $m$ is equal to zero; when measured in the Fourier basis, the measurement outcomes are all the same. 

The other $n$-level $n$-particle state $\ket{S_n}$ in computational basis is described as:
\begin{align}
	\ket{S_n}_c = \frac{1}{\sqrt{n!}} \sum_{S \in P_n^n} (-1)^{\tau(S)} \ket{s_0}_c \ket{s_1}_c \cdots \ket{s_{n-1}}_c
\end{align}
where $P_n^n$ represents the set of all permutations of $\{0,1,\cdots,n-1\}$, $S$ is a permutation in the form $S=\{s_0,s_1,\cdots,s_{n-1}\}$ and $\tau(S)$ is defined as the number of transpositions of pairs of elements of S that must be composed to place the elements in canonical order $012\cdots{n-1}$ .

$\ket{S_n}$ has the property that the permutation of the outcomes of n particles is a random element of the set $P_n^n$, regardless of whether it is measured in the computational basis or in the Fourier basis.

These two special quantum entangled states will be used to construct the ballot matrix and ballot index respectively, and their special properties will be used for security test.

\subsection{Hybrid Network and Role Model}

\begin{figure}[b]
	\centering
	\includegraphics[width=0.5\textwidth]{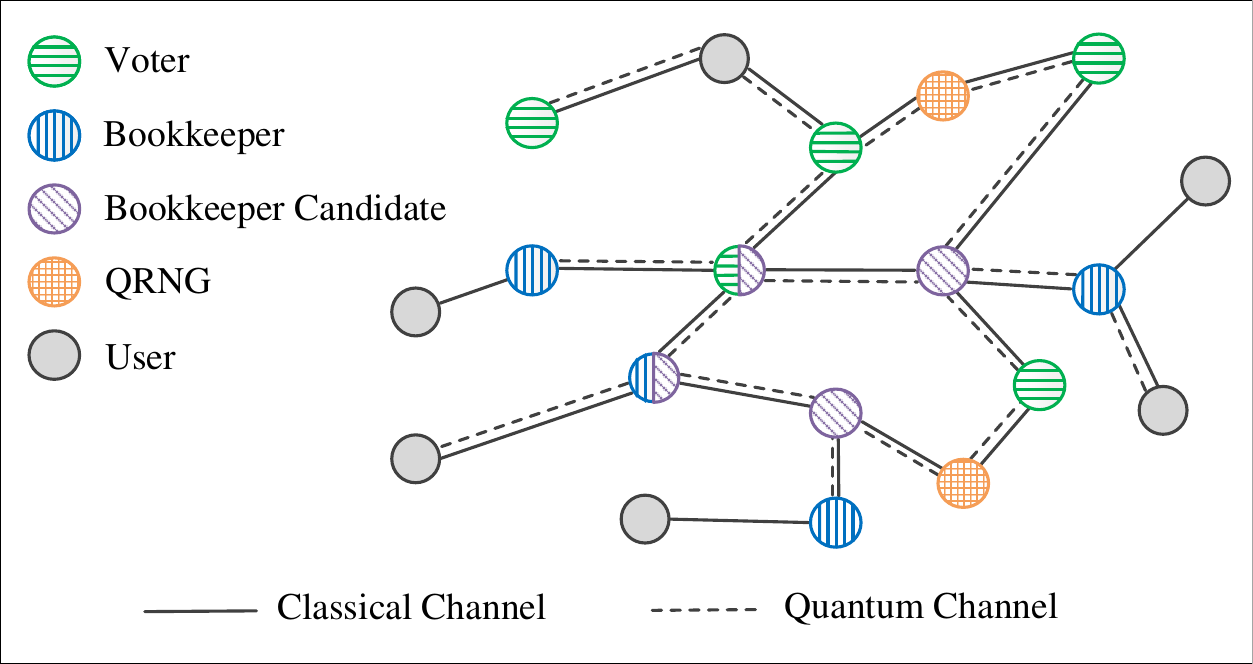}
	\caption{A possible hybrid network model with five different roles.} 
	\label{Fig5}
\end{figure}

Q-PnV adopts a hybrid network model that integrates classical and quantum networks. The hybrid network includes five types of nodes: voters, bookkeepers, bookkeeper candidates, ordinary users, and QRNGs. Some nodes may simultaneously serve two roles. Fig.5 illustrates a possible network model. Voters, bookkeepers, bookkeeper candidates, and QRNGs must connect to the quantum network, while ordinary users are not required to connect if they do not intend to participate in the consensus process. The responsibilities of each role are described as follows:

\textbf{Voters}: Voters are responsible for block validation and the election of bookkeepers, both of which are conducted through voting. For each block in the network, voters review it and decide whether it is valid. Only blocks that receive more than $50\%$ approval are considered valid and added to the blockchain. At the end of each tenure cycle, voters evaluate the bookkeeper candidates and vote to determine if they will become the bookkeepers in the next tenure cycle.

\textbf{Bookkeepers}: There are multiple bookkeepers in the network, and all bookkeepers in a tenure cycle produce blocks in a random order. The rotating bookkeeper responsible for block production collects relevant transactions from the network, assembles them into blocks, and broadcasts them across the network for validation. The validation process requires bookkeepers to prepare and distribute the corresponding quantum states.

\textbf{Bookkeeper Candidates}: Bookkeeper candidates do not directly participate in the consensus process. At the end of each tenure cycle, all bookkeepers automatically become bookkeeper candidates and participate in the election of bookkeepers for the next tenure cycle. Any node has the right to apply to become a bookkeeper candidate, and candidates can voluntarily withdraw from the process of becoming bookkeepers.

\textbf{Ordinary Users}: Ordinary users do not have any responsibilities in the consensus process. They can only receive verified data from other nodes, and their joining or leaving does not significantly affect the overall network dynamics.

\textbf{QRNG}: QRNGs are device nodes within the network, jointly maintained by the consortium. They are responsible for generating random numbers during the consensus process.

\begin{figure}[b]
	\centering
	\includegraphics[width=0.5\textwidth]{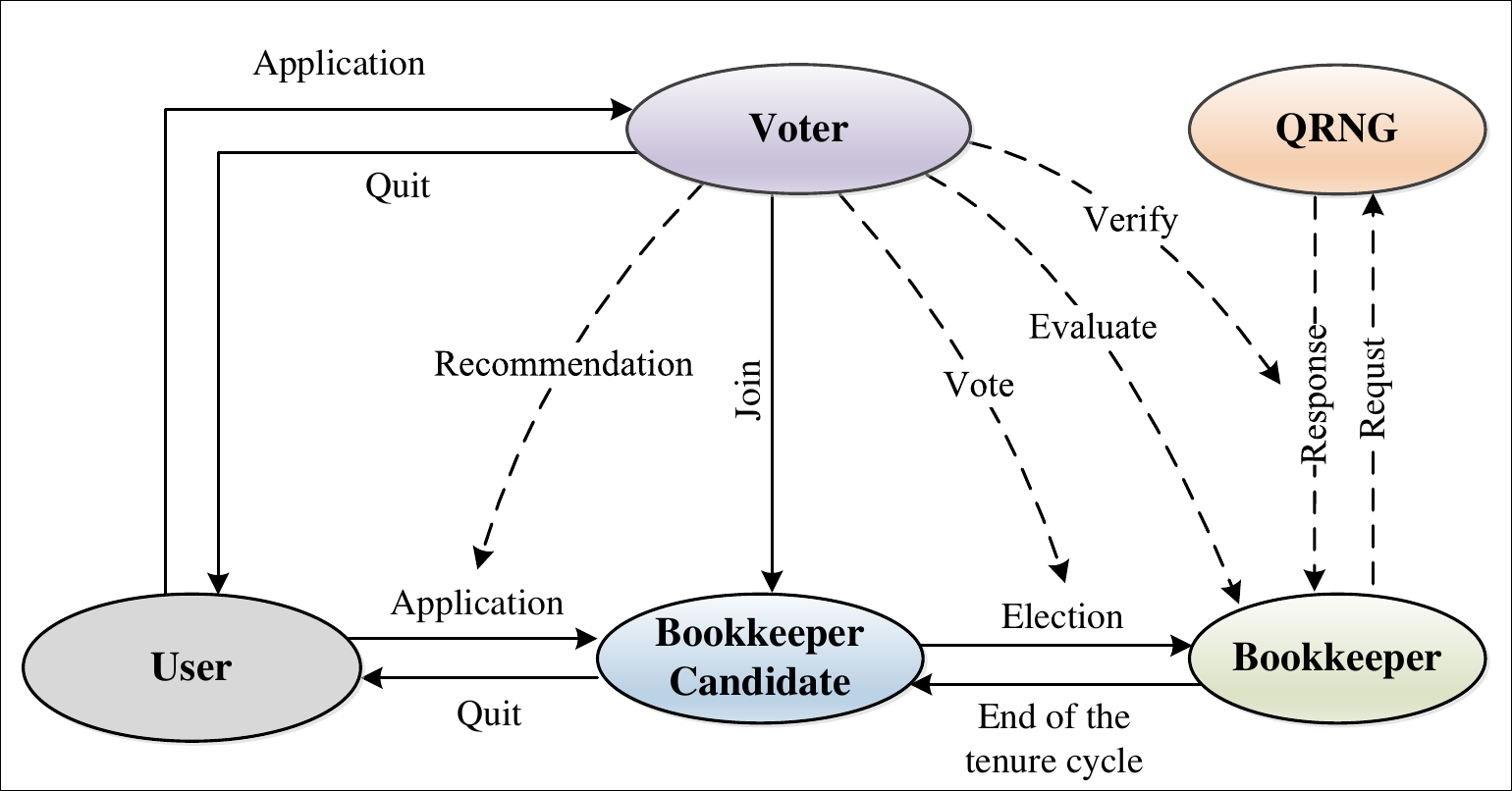}
	\caption{The transition relationship among different roles in the network.} 
	\label{Fig6}
\end{figure}

The transition relationships among different roles are illustrated in Fig.6. QRNGs function solely as device nodes within the network and do not participate in role transitions. Notably, ordinary users who wish to become voters or bookkeeper candidates must connect to the quantum network. Additionally, since voters need the capability to prepare quantum entangled states, only those voters or ordinary users with the necessary equipment can apply to become bookkeeper candidates.

In a consortium blockchain, all nodes participating in the consensus process must undergo identity authentication. However, classical public-private key methods face security risks in the presence of quantum computing, making classical identity authentication potentially vulnerable to forgery. Quantum identity authentication ensures absolute security in this process. Q-PnV employs an identity authentication method based on QKD.

Ordinary users wishing to become voters or bookkeeper candidates need to undergo identity authentication. The user must first share a quantum key with the consortium committee through a QKD protocol. This quantum key ensures the absolute security of the information. The user then sends encrypted identity information and a hash tag generated using this key to the committee. Upon receiving the message, the committee recalculates the hash tag. If the two hash tags match, the sender’s identity is verified. The use of QKD-based identity authentication naturally excludes users not connected to the quantum network, eliminating the need to verify users’ capability to transmit quantum information.

\subsection{Quantum Voting in Q-PnV}
Quantum Voting is the core process of Q-PnV, where voters determine the validation of blocks produced by bookkeepers and elect bookkeepers in the next tenure cycle. Q-PnV employs a voting protocol \cite{wang_self-tallying_2016} based on the two special quantum multipartite entangled states $\ket{X_n}$ and $\ket{S_n}$.

In the voting on block validation, the results are classified as either valid or invalid, so the level of $\ket{X_n}$ and $\ket{S_n}$ equals $2$. The voting process follows three steps, as shown in Fig.7.

\begin{figure}[b]
	\centering
	\includegraphics[width=0.45\textwidth]{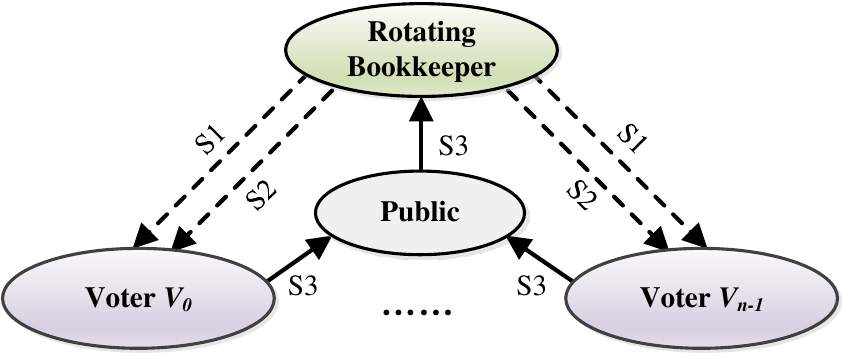}
	\caption{Three steps of quantum voting in Q-PnV.} 
	\label{Fig7}
\end{figure}

\textbf{S1 Distribute ballot boxes}

The rotating bookkeeper prepares $n+n\delta_0$ copies of $2$-level entangled states $\ket{X_n}$, where $\delta_0$ is the security strength of ballot boxes and $n$ represents the number of voters. The $x$-th copy of $\ket{X_n}$ consists of $n$ particles $p_{x,0},p_{x,1},\cdots,p_{x,n-1}$, and all particles form a particle matrix $p_{(n+n\delta_0)\times n}$ with $n+n\delta_0$ rows and $n$ columns. Subsequently, the rotating bookkeeper distributes all the columns of this matrix to the corresponding voters, where the $k$-th column $S_k=\{p_{0,k},p_{1,k},\cdots,p_{n+n\delta_0-1,k}\}^T$ is distributed to voter $V_k$. Here, $p_{x,k}$ denotes the particle in row $x$ and column $k$ of the particle matrix, $0\leq x\leq n+n\delta_0-1, 0\leq k\leq n-1$.

After the preparing and distributing, each voter will perform a security test to ensure that the distributed quantum states are intact. Since the order of test does not affect the results, we assume that voter $V_0$ starts the test. $V_0$ randomly picks out $\delta_0$ particles as the test particles, and chooses from computational or Fourier basis with uniform distribution for each test particle. Then the row index of test particles $\{x_0,x_1,\cdots,x_{\delta_0-1}\}$ and the selected measurement basis are published. After receiving this public information, all other voters are required to measure their test particles of the same row index with the same measurement basis. The measurement result of a test particle in row $x_l$ and column $k$ is denoted by $r_{x_l,k}$. The properties of $\ket{X_n}$ are used to check whether the quantum state is intact. If $V_0$ chooses the computational basis, it needs to check whether $\sum_{l=0}^{n-1} r_{{x_l},k} \mod 2 =0 $ holds true. If $V_0$ chooses the Fourier basis, it must verify whether $r_{x_l,0},r_{x_l,1},\cdots,r_{x_l,n-1}$ are identical. 

The failure of the above condition indicates a test failure, requiring $V_0$ to notify other nodes to abort the protocol. Conversely, the satisfaction of either condition signifies a test success, in which case the same testing procedure is executed by the subsequent voters. Repeat the same procedure until the test performed by each voter is passed or abort the protocol in some intermediate process.

After each security test, $\delta_0$ test particles are discarded. After all the voters have passed the security test, each voter retains only n valid particles, and uses the computational basis to measure its particles to obtain n measurement values. We use a matrix $r_{n \times n}$ with n rows and n columns to denote the measurement results of all voters. The k-th column of matrix $r$ corresponds to the measurement results of particles held by $V_k$, which also represents the ballot of the voter $V_k$. Through subsequent renewal of the matrix, the voters can anonymously convey voting information, so we refer to matrix $r$ as the ballot matrix.

\textbf{S2 Distribute ballot indexes}

Similar to the $\ket{X_n}$, the rotating bookkeeper prepares $1+n\delta_1$ copies of the entangled state $\ket{S_n}$ and distribute them to corresponding voters, where $n\delta_1$ is the security strength of ballot indexes. A security test is then conducted to ensure the integrity of $\ket{S_n}$. $n\delta_1$ particles are randomly selected as test particles, and computational basis or Fourier basis is also randomly selected as the measurement basis to measure the test particles according to the published row index. Also, the property of $\ket{S_n}$ is used to check whether the quantum state is intact.

After the security tests pass, all voters discard the test particles. Each voter subsequently has only one valid particle left and measures the particle on the computational basis. Suppose that the valid particles are $t_{0,0},t_{0,1},\cdots,t_{0,n-1}$, the measurement result $d_{0,k}$ corresponds to an index used to denote the ballot box selection of the voter $V_k$.

\textbf{S3 Vote casting }

After S1 and S2, each voter $V_k$ now has $n$ ballot numbers $r_{0,k},r_{1,k},\cdots,r_{n-1,k}$ and one index number $d_{0,k}$. Suppose the voting option of the voter $V_k$ is $v_k$, $r'_{x,k}$ denotes the modified $r_{x,k}$, voter $V_k$ votes to $v_k \in \{0,1\}$ by adding $v_k$ to $r_{x,k}$:
\begin{align}
	r'_{x,k} =
	\begin{cases}
		r_{x,k} + v_k \mod 2, & \text{if } x = d_{0,k}, \\
		r_{x,k}, & \text{if } x \neq d_{0,k}.
	\end{cases}
\end{align}

All voters publish their updated votes to obtain the final ballot matrix $r'$. By summing $r'$ by row, the number of votes received by the $l$-th voting item can be calculated as follows:

\begin{align}
	R_x = \sum_{k=0}^{n-1} r'_{x,k} \mod 2
\end{align}

\begin{align}
	N_l = \sum_{R_x = l} 1
\end{align}
where $R_x$ denotes the summation result of the $x$-th row of the updated ballot matrix $r'$ and $N_l$ is used to count the number of occurrences of $R_x=l$. We use $0$ to denote the block verification passes, and $1$ to denote the verification fails. When $N_0>n/2$, meaning more than half of the voters have voted to pass, the block verification for that round is completed. After the block verification is completed, the rotating bookkeeper writes the voting results into the block and releases the official block to the entire network. This quantum voting protocol has the characteristic of self-tally, allowing voters to independently verify the accuracy of the voting results.

The bookkeepers for the next tenure cycle are also elected through this voting process. Voters will decide whether to nominate the corresponding bookkeeper candidates for the next tenure cycle. A vote of $0$ indicates a recommendation, while a vote of $1$ indicates a rejection. The final results are ranked based on the number of votes received, and the top $N_B$ candidates will become bookkeepers, where $N_B$ represents the number of bookkeepers in each tenure cycle.

\subsection{Determining the Rotating Bookkeeper}
In PoV, a random number is used to select the next rotating bookkeeper. The current bookkeeper generates the random number through a deterministic algorithm that relies on the modulus operation of hashed signatures and timestamps. This random number generation algorithm is not truly random and cannot cope with the increasing quantum threats. In Q-PnV, QRNG is used to generate random numbers to determine the next rotating bookkeeper. Considering the cost issue of QRNG, Q-PnV sets QRNG as one or more nodes in the network. In a consortium blockchain, the availability of QRNG is maintained and supervised jointly by the entire consortium, ensuring it is a true quantum device and does not behave dishonestly. Therefore, it is only necessary to consider the behavior of other nodes impersonating QRNG, which can be resolved through quantum identity authentication.

When block generation is confirmed, the rotating bookkeeper will communicate with the QRNG and request a random number to determine the next rotating bookkeeper. Upon receiving the request, the QRNG will generate a random number, send it to the bookkeeper, and keep a local copy for authentication. The rotating bookkeeper will write this number into the block and send it to the entire network.

To ensure that the random number is generated by the QRNG, voters can verify this process. This requires each voter to share a quantum key with the QRNG. When a voter sends a verification request, the QRNG will send the random number and the hash tag generated using the corresponding key to the voter. Upon receiving the message, the voter will recalculate the hash tag, and if both match, the identity can be verified, ensuring that the random number is generated by the QRNG. This QRNG-based random number generation method ensures the security and randomness of the rotation order, eliminating the need to rely on the quantity relationship between different roles for security as in PoV.

Mannalatha et al. reviewed and summarized all current QRNGs \cite{mannalatha_comprehensive_2023}, classified the available QRNGs, and rigorously analyzed the technical challenges associated with each category. The choice of device depends entirely on the actual needs of the consortium.

\section{Quantum Consortium Blockchain based on Q-PnV}
Combining Q-PnV, which is suitable for consortium blockchains, with quantum blockchain schemes results in a quantum consortium blockchain. Among the currently proposed fully quantum blockchain schemes, we adopt a quantum blockchain based on weighted hypergraphs due to its concrete implementation method. This scheme has been utilized in multiple quantum blockchain protocols \cite{banerjee_quantum_2020, li_efficient_2022}.

\subsection{Determining the Rotating Bookkeeper}
The construction of a blockchain based on Q-PnV involves three steps.

\textbf{S1 Construction of the quantum block} 

The quantum block needs to encode the classical information $p$ into the phase of a qubit, with one qubit corresponding to one block. The rotating bookkeeper prepares the initial qubit $\ket{+}$, and then introduces $p$ into the qubit. In addition to the qubits, blocks also contain the next rotating bookkeeper and other necessary data.
After the block is constructed, the current rotating bookkeeper sends the block and information to all voters. Since the rotating bookkeeper knows the qubit information precisely, it is feasible to prepare identical copies of the qubit without violating the no-cloning theorem.

\textbf{S2 Implementing the Q-PnV consensus mechanism}

After receiving the block, voting will be conducted according to the aforementioned Q-PnV consensus mechanism. Voters need to assess whether the transactions are valid, determine if the block satisfies $\theta_p\in(0,{\pi}/2)$, $\sum_{i}\theta_{p_i}<{\pi}/2$ and verify if the next rotating bookkeeper is correctly selected. If all criteria are met, they can vote to approve the block generation. The current rotating bookkeeper counts the votes, and if more than half of the voters agree, the block is considered valid.

\textbf{S3 Adding the quantum block to the entanglement chain}

If the block is valid, nodes use quantum entanglement operations to add the block to the local quantum blockchain using Controlled-Z gates. For the entanglement of two blocks, one Controlled-Z gate can be used directly. However, for three or more blocks, a Controlled-Z gate with multiple control qubits does not currently exist, so the quantum circuit of a weighted hypergraph cannot be used directly. In the construction of three blocks, Banerjee et al. used a Toffoli gate acting on auxiliary qubits \cite{banerjee_quantum_2020}. Li et al. proposed a quantum circuit using weighted graph states that only requires Controlled-Z gates with a single control qubit \cite{li_efficient_2022}.

The drawback is that their circuit requires more non-Clifford+T gates. The Clifford+T gate set is a universal set, meaning any quantum circuit can be approximated using only these types of gates. This gate set is compatible with error detection and correction codes, effectively combating noise issues. Orts et al. further optimized Banerjee's approach \cite{orts_improving_2023}, improving the quantum circuit diagram to significantly reduce the number of quantum gates and circuit depth, and provided quantum circuit diagrams for $4$-block and $5$-block configurations. In this paper, we focus on the results of the entanglement rather than the process. After an effective quantum circuit, the quantum block successfully achieves entanglement.

It should be noted that Q-PnV divides blocks into ordinary blocks and special blocks, similar to PoV. However, only the internal data of the blocks differ, and the construction method of the blocks is the same. Fig.8 shows all the blocks generated during one tenure cycle, including n ordinary blocks and one special block, linked by entanglement.

Affected by the entanglement process, a quantum blockchain using weighted hypergraphs cannot resolve situations where two or more peers send their blocks simultaneously. However, a consortium blockchain based on Q-PnV naturally avoids this issue. In the absence of network partitions, there will only be one rotating bookkeeper at any given time, so only one block can be generated. In the case of network partitions, since the quantum voting results of the partition with fewer voters will fall below the required threshold, at most one block will be generated.

\begin{figure*}[t]
	\centering
	\includegraphics[width=0.75\textwidth]{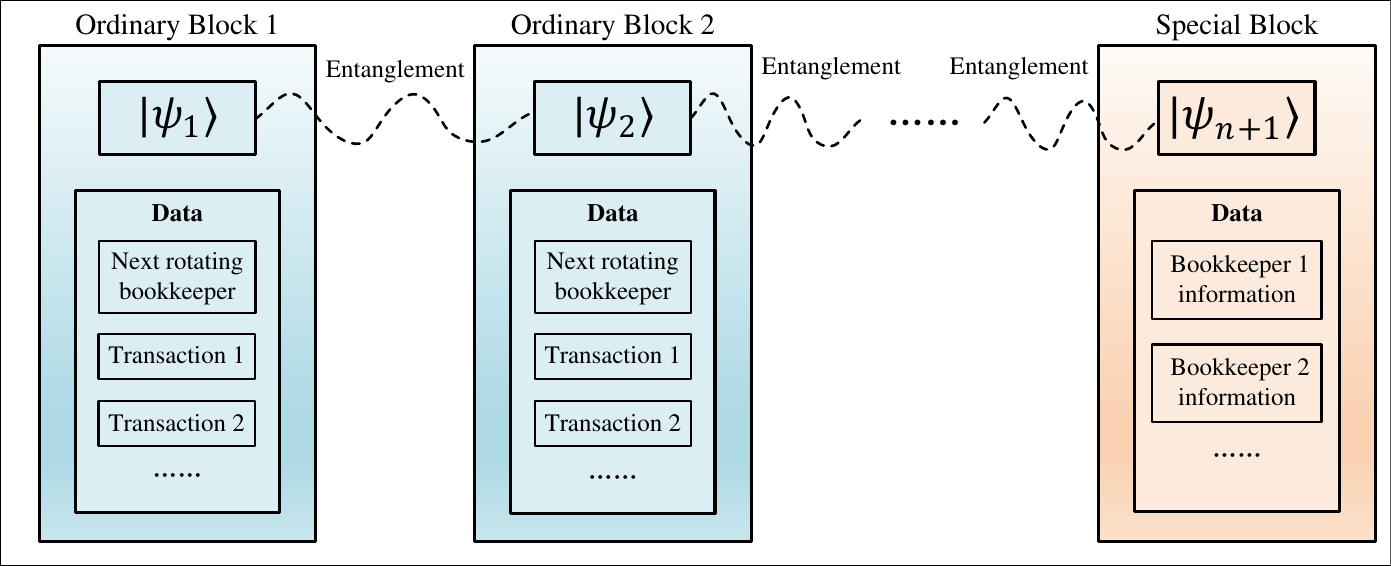}
	\caption{Quantum block generated during one tenure cycle are linked by entanglement.} 
	\label{Fig8}
\end{figure*}

\subsection{Example of the Quantum Consortium Blockchain}

Here is an example of constructing a quantum blockchain using Q-PnV. Assuming there are $4$ voters, and the rotating bookkeeper has already been selected. Suppose the rotating bookkeeper here constructs a ordinary block. The consortium sets the security strength value as $\delta_0=\delta_1=1$.

The rotating bookkeeper distributes the quantum state $\ket{X_n}$, and all voters measure their particles in the computational basis. The ballot matrix $r$ is assumed to be:
\begin{align}
	r = \begin{pmatrix}
		0 & 1 & 0 & 1 \\
		1 & 1 & 0 & 0 \\
		1 & 0 & 0 & 1 \\
		0 & 0 & 1 & 1 \\
		1 & 1 & 1 & 1 \\
		1 & 0 & 1 & 0 \\
		0 & 1 & 1 & 0 \\
		0 & 0 & 0 & 0
	\end{pmatrix}
\end{align}

Suppose four voters $V_0,V_1,V_2,V_3$ select indexes of test particles as $4, 5, 6, 7$ respectively. $V_0$ verifies $(1+1+1+1)\mod 2=0$, $V_1$ verifies $(1+0+1+0)\mod 2=0$, $V_2$ verifies $(0+1+1+0)\mod 2=0$ and $V_3$ verifies $(0+0+0+0)\mod 2=0$. After the security test, the test particle are discarded, resulting in the final ballot matrix:
\begin{align}
	r = \begin{pmatrix}
		0 & 1 & 0 & 1 \\
		1 & 1 & 0 & 0 \\
		1 & 0 & 0 & 1 \\
		0 & 0 & 1 & 1 
	\end{pmatrix}
\end{align}

The rotating bookkeeper distributes the quantum state $\ket{S_n}$, and all voters measure their particles in the computational basis. The matrix $d$ is assumed to be:
\begin{align}
	d = \begin{pmatrix}
		0 & 1 & 3 & 2 \\
		1 & 2 & 3 & 0 \\
		1 & 0 & 2 & 3 \\
		3 & 1 & 0 & 2 \\
		0 & 3 & 2 & 1
	\end{pmatrix}
\end{align}

In the second security test, suppose four voters $V_0,V_1,V_2,V_3$ select indexes of test particles as $1, 2, 3, 4$. $V_0$ verifies $\{1,2,3,0\}\in P_n^n$, $V_1$ verifies $\{1,0,2,3\}\in P_n^n$, $V_2$ verifies $\{0,3,2,1\}\in P_n^n$ and $V_3$ verifies $\{0,3,2,1\}\in P_n^n$. After the test, the test particle are discarded, resulting in the final ballot index sequence:
\begin{align}
	d = (0, 1, 3, 2)
\end{align}

Assume that voting sequence of all voters is $v=\{0,0,1,0\}$, denoting $V_0$, $V_1$ and $V_3$ pass the verification. The process of voting statistics is shown in Table \uppercase\expandafter{\romannumeral1}. The final voting result sequence is $R=\{0,0,0,1\}$, $N_0=\sum_{R_x=0}1=3>n/2$ and the block verification passed in this rotation round.

Suppose three blocks were generated using the aforementioned method, with phase angles $\theta_{p_1}=\pi/8$, $\theta_{p_2}=\pi/16$ and $\theta_{p_3} =\pi/32$ corresponding to the quantum states $\ket{\psi_1}$, $\ket{\psi_2}$ and $\ket{\psi_3}$. Then, a quantum circuit is used to entangle the three quantum states. The improved quantum circuit shown in the Fig. 9 can be used for this purpose. Finally, the quantum blockchain based on Q-PnV is successfully formed.

\begin{figure}[t]
	\centering
	\includegraphics[width=0.40\textwidth]{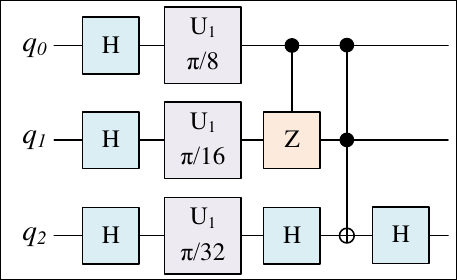}
	\caption{Implementation of a 3-blockchain circuit proposed by Orts et al. \cite{orts_improving_2023}.} 
	\label{Fig9}
\end{figure}

\begin{table}[ht]
	\renewcommand\arraystretch{1.5}
	\centering
	\caption{The Process of Voting Statistics}
	\begin{tabular}{m{1cm}<{\centering} | m{1cm}<{\centering} m{1cm}<{\centering} m{1cm}<{\centering} m{1.2cm}<{\centering} m{1.2cm}<{\centering}}
		\hline
		& \textbf{$V_0$} & \textbf{$V_1$} & \textbf{$V_2$} & \textbf{$V_3$} & \textbf{$R$}  \\ \hline
		\textbf{$r'_{0,k}$} & 0+0   & 1     & 0     & 1     & 0   \\ 
		\textbf{$r'_{1,k}$} & 1     & 1+0   & 0     & 0     & 0   \\ 
		\textbf{$r'_{2,k}$} & 1     & 0     & 0     & 1+0   & 0   \\ 
		\textbf{$r'_{3,k}$ }& 0     & 0     & 1+1   & 1     & 1   \\ \hline
	\end{tabular}
\end{table}

\section{Analysis}
\subsection{Security}
The quantum blockchain based on Q-PnV is adaptable to the future quantum era, as attackers, regardless of their quantum computing power, cannot tamper with the blocks or attack the consensus process. Since the blocks are linked through entanglement, any modification to a block will disrupt the entire chain. Thanks to the self-counting and fairness advantages of quantum voting, attackers cannot interfere with the voting process. Voters can perform security tests before voting to ensure the safety of the quantum state and can also self-tally votes afterward to verify the impartiality of the rotating bookkeeper.

\subsection{Fairness}
In the consortium blockchain, block generation is handled by bookkeepers, and the selection of bookkeepers is collectively decided by the voters. The use of QRNG to generate random numbers during the consensus process ensures that the selection of the rotating bookkeeper is completely random each time, preventing the situation where one or a few bookkeepers control block production. Therefore, as long as voters adopt an appropriate method for selecting bookkeepers, such as scoring candidates to determine their suitability, the block production process will be entirely fair.

\section{Conclusion}
This paper proposes a novel quantum consensus mechanism Q-PnV, which is suitable for consortium blockchains. This consensus mechanism can be combined with existing quantum blockchain schemes to achieve a quantum blockchain in consortium blockchain scenarios. This fully quantum consortium blockchain is expected to ensure security and fairness in the future quantum era. However, due to technical limitations, the proposed scheme cannot yet be experimented with or simulated.

\section*{Acknowledgments}
This work is supported by GuangDong Prov., R\&D Key Program (No.2019B010137001), Peking University Shenzhen Graduate School - China Mobile Internet Co., Ltd. Joint Laboratory for Sovereign and Trusted Internet, Basic Research Enhancement Program of China under Grant (2021-JCJQ-JJ-0483), GuangDong Prov. Basic Research (No. 2022A1515010836), China Environment for Network Innovation (CENI) (GJFGW No.[2020]386, SZFGW [2019]261), Shenzhen Research Programs (JCYJ20220531093206015, JCYJ20210324122013036, JCYJ20190808155607340), Shenzhen Fundamental Research Program (No.GXWD20201231165807007-20200807164903001), National Keystone R\&D Program of China (No. 2017YFB0803204), NSFC (No. 61671001), ZTE Funding (No.2019ZTE03-01), HuaWei Funding (No.TC20201222002).

\bibliography{References}

\begin{thebibliography}{10}
\providecommand{\url}[1]{#1}
\csname url@samestyle\endcsname
\providecommand{\newblock}{\relax}
\providecommand{\bibinfo}[2]{#2}
\providecommand{\BIBentrySTDinterwordspacing}{\spaceskip=0pt\relax}
\providecommand{\BIBentryALTinterwordstretchfactor}{4}
\providecommand{\BIBentryALTinterwordspacing}{\spaceskip=\fontdimen2\font plus
\BIBentryALTinterwordstretchfactor\fontdimen3\font minus \fontdimen4\font\relax}
\providecommand{\BIBforeignlanguage}[2]{{%
\expandafter\ifx\csname l@#1\endcsname\relax
\typeout{** WARNING: IEEEtran.bst: No hyphenation pattern has been}%
\typeout{** loaded for the language `#1'. Using the pattern for}%
\typeout{** the default language instead.}%
\else
\language=\csname l@#1\endcsname
\fi
#2}}
\providecommand{\BIBdecl}{\relax}
\BIBdecl

\bibitem{shor_algorithms_1994}
\BIBentryALTinterwordspacing
P.~Shor, ``Algorithms for quantum computation: discrete logarithms and factoring,'' in \emph{Proceedings 35th {Annual} {Symposium} on {Foundations} of {Computer} {Science}}.\hskip 1em plus 0.5em minus 0.4em\relax Santa Fe, NM, USA: IEEE Comput. Soc. Press, 1994, pp. 124--134. [Online]. Available: \url{http://ieeexplore.ieee.org/document/365700/}
\BIBentrySTDinterwordspacing

\bibitem{grover_quantum_1997}
\BIBentryALTinterwordspacing
L.~K. Grover, ``\BIBforeignlanguage{en}{Quantum {Mechanics} {Helps} in {Searching} for a {Needle} in a {Haystack}},'' \emph{\BIBforeignlanguage{en}{Physical Review Letters}}, vol.~79, no.~2, pp. 325--328, Jul. 1997. [Online]. Available: \url{https://link.aps.org/doi/10.1103/PhysRevLett.79.325}
\BIBentrySTDinterwordspacing

\bibitem{yang_survey_2024}
\BIBentryALTinterwordspacing
Z.~Yang, H.~Alfauri, B.~Farkiani, R.~Jain, R.~D. Pietro, and A.~Erbad, ``A {Survey} and {Comparison} of {Post}-{Quantum} and {Quantum} {Blockchains},'' \emph{IEEE Communications Surveys \& Tutorials}, vol.~26, no.~2, pp. 967--1002, 2024. [Online]. Available: \url{https://ieeexplore.ieee.org/document/10288193/}
\BIBentrySTDinterwordspacing

\bibitem{cao_evolution_2022}
\BIBentryALTinterwordspacing
Y.~Cao, Y.~Zhao, Q.~Wang, J.~Zhang, S.~X. Ng, and L.~Hanzo, ``The {Evolution} of {Quantum} {Key} {Distribution} {Networks}: {On} the {Road} to the {Qinternet},'' \emph{IEEE Communications Surveys \& Tutorials}, vol.~24, no.~2, pp. 839--894, 2022. [Online]. Available: \url{https://ieeexplore.ieee.org/document/9684555/}
\BIBentrySTDinterwordspacing

\bibitem{kiktenko_quantum-secured_2018}
\BIBentryALTinterwordspacing
E.~O. Kiktenko, N.~O. Pozhar, M.~N. Anufriev, A.~S. Trushechkin, R.~R. Yunusov, Y.~V. Kurochkin, A.~I. Lvovsky, and A.~K. Fedorov, ``Quantum-secured blockchain,'' \emph{Quantum Science and Technology}, vol.~3, no.~3, p. 035004, Jul. 2018. [Online]. Available: \url{https://iopscience.iop.org/article/10.1088/2058-9565/aabc6b}
\BIBentrySTDinterwordspacing

\bibitem{rajan_quantum_2019}
\BIBentryALTinterwordspacing
D.~Rajan and M.~Visser, ``\BIBforeignlanguage{en}{Quantum {Blockchain} {Using} {Entanglement} in {Time}},'' \emph{\BIBforeignlanguage{en}{Quantum Reports}}, vol.~1, no.~1, pp. 3--11, Apr. 2019. [Online]. Available: \url{https://www.mdpi.com/2624-960X/1/1/2}
\BIBentrySTDinterwordspacing

\bibitem{banerjee_quantum_2020}
\BIBentryALTinterwordspacing
S.~Banerjee, A.~Mukherjee, and P.~K. Panigrahi, ``\BIBforeignlanguage{en}{Quantum blockchain using weighted hypergraph states},'' \emph{\BIBforeignlanguage{en}{Physical Review Research}}, vol.~2, no.~1, p. 013322, Mar. 2020. [Online]. Available: \url{https://link.aps.org/doi/10.1103/PhysRevResearch.2.013322}
\BIBentrySTDinterwordspacing

\bibitem{li_efficient_2022}
\BIBentryALTinterwordspacing
Q.~Li, J.~Wu, J.~Quan, J.~Shi, and S.~Zhang, ``Efficient {Quantum} {Blockchain} {With} a {Consensus} {Mechanism} {QDPoS},'' \emph{IEEE Transactions on Information Forensics and Security}, vol.~17, pp. 3264--3276, 2022. [Online]. Available: \url{https://ieeexplore.ieee.org/document/9872051/}
\BIBentrySTDinterwordspacing

\bibitem{orts_improving_2023}
\BIBentryALTinterwordspacing
F.~Orts, R.~Paulavičius, and E.~Filatovas, ``\BIBforeignlanguage{en}{Improving the implementation of quantum blockchain based on hypergraphs},'' \emph{\BIBforeignlanguage{en}{Quantum Information Processing}}, vol.~22, no.~9, p. 330, Sep. 2023. [Online]. Available: \url{https://link.springer.com/10.1007/s11128-023-04096-w}
\BIBentrySTDinterwordspacing

\bibitem{nilesh_quantum_2022}
\BIBentryALTinterwordspacing
K.~Nilesh and P.~K. Panigrahi, ``Quantum {Blockchain} {Based} on {Dimensional} {Lifting} {Generalized} {Gram}-{Schmidt} {Procedure},'' \emph{IEEE Access}, vol.~10, pp. 103\,212--103\,222, 2022. [Online]. Available: \url{https://ieeexplore.ieee.org/document/9895383/}
\BIBentrySTDinterwordspacing

\bibitem{wen_blockchain_2022}
\BIBentryALTinterwordspacing
X.-J. Wen, Y.-Z. Chen, X.-C. Fan, W.~Zhang, Z.-Z. Yi, and J.-B. Fang, ``\BIBforeignlanguage{en}{Blockchain consensus mechanism based on quantum zero-knowledge proof},'' \emph{\BIBforeignlanguage{en}{Optics \& Laser Technology}}, vol. 147, p. 107693, Mar. 2022. [Online]. Available: \url{https://linkinghub.elsevier.com/retrieve/pii/S0030399221007817}
\BIBentrySTDinterwordspacing

\bibitem{wen_blockchain_2022-1}
\BIBentryALTinterwordspacing
X.~Wen, Y.~Chen, W.~Zhang, Z.~L. Jiang, and J.~Fang, ``\BIBforeignlanguage{en}{Blockchain {Consensus} {Mechanism} {Based} on {Quantum} {Teleportation}},'' \emph{\BIBforeignlanguage{en}{Mathematics}}, vol.~10, no.~14, p. 2385, Jul. 2022. [Online]. Available: \url{https://www.mdpi.com/2227-7390/10/14/2385}
\BIBentrySTDinterwordspacing

\bibitem{wang_consensus_2022}
\BIBentryALTinterwordspacing
P.~Wang, W.~Chen, S.~Lin, L.~Liu, Z.~Sun, and F.~Zhang, ``\BIBforeignlanguage{en}{Consensus algorithm based on verifiable quantum random numbers},'' \emph{\BIBforeignlanguage{en}{International Journal of Intelligent Systems}}, vol.~37, no.~10, pp. 6857--6876, Oct. 2022. [Online]. Available: \url{https://onlinelibrary.wiley.com/doi/10.1002/int.22865}
\BIBentrySTDinterwordspacing

\bibitem{li_proof_2017}
\BIBentryALTinterwordspacing
K.~Li, H.~Li, H.~Hou, K.~Li, and Y.~Chen, ``Proof of {Vote}: {A} {High}-{Performance} {Consensus} {Protocol} {Based} on {Vote} {Mechanism} \& {Consortium} {Blockchain},'' in \emph{2017 {IEEE} 19th {International} {Conference} on {High} {Performance} {Computing} and {Communications}; {IEEE} 15th {International} {Conference} on {Smart} {City}; {IEEE} 3rd {International} {Conference} on {Data} {Science} and {Systems} ({HPCC}/{SmartCity}/{DSS})}.\hskip 1em plus 0.5em minus 0.4em\relax Bangkok: IEEE, Dec. 2017, pp. 466--473. [Online]. Available: \url{http://ieeexplore.ieee.org/document/8291964/}
\BIBentrySTDinterwordspacing

\bibitem{li_pov_2020}
\BIBentryALTinterwordspacing
K.~Li, H.~Li, H.~Wang, H.~An, P.~Lu, P.~Yi, and F.~Zhu, ``{PoV}: {An} {Efficient} {Voting}-{Based} {Consensus} {Algorithm} for {Consortium} {Blockchains},'' \emph{Frontiers in Blockchain}, vol.~3, p.~11, Mar. 2020. [Online]. Available: \url{https://www.frontiersin.org/article/10.3389/fbloc.2020.00011/full}
\BIBentrySTDinterwordspacing

\bibitem{bai_parallel_2021}
\BIBentryALTinterwordspacing
Y.~Bai, Y.~Zhi, H.~Li, H.~Wang, P.~Lu, and C.~Ma, ``\BIBforeignlanguage{en}{On {Parallel} {Mechanism} of {Consortium} {Blockchain}: {Take} {PoV} as an example},'' in \emph{\BIBforeignlanguage{en}{2021 {The} 3rd {International} {Conference} on {Blockchain} {Technology}}}.\hskip 1em plus 0.5em minus 0.4em\relax Shanghai China: ACM, Mar. 2021, pp. 147--154. [Online]. Available: \url{https://dl.acm.org/doi/10.1145/3460537.3460560}
\BIBentrySTDinterwordspacing

\bibitem{wang_pnv_2024}
\BIBentryALTinterwordspacing
H.~Wang, H.~Li, P.~Fan, J.~Kang, S.~Deng, and X.~Zhu, ``\BIBforeignlanguage{en}{{PnV}: {An} {Efficient} {Parallel} {Consensus} {Protocol} {Integrating} {Proof} and {Voting}},'' \emph{\BIBforeignlanguage{en}{Applied Sciences}}, vol.~14, no.~8, p. 3510, Apr. 2024. [Online]. Available: \url{https://www.mdpi.com/2076-3417/14/8/3510}
\BIBentrySTDinterwordspacing

\bibitem{li_CAP_2024}
H.~Wang and H.~Li, ``Principles and {Applications} of {Blockchain} {Systems}: {How} to {Break} {Through} {Consensus} {Trilemma} in {Consortium} {Blockchains},'' in \emph{Wiley-IEEE {Press}}, 2024, pp. 72--76, unpublished.

\bibitem{castro_practical_1999}
M.~Castro, B.~Liskov, and {others}, ``Practical byzantine fault tolerance,'' in \emph{{OsDI}}, vol.~99, 1999, pp. 173--186, issue: 1999.

\bibitem{sukhwani_performance_2017}
\BIBentryALTinterwordspacing
H.~Sukhwani, J.~M. Martinez, X.~Chang, K.~S. Trivedi, and A.~Rindos, ``Performance {Modeling} of {PBFT} {Consensus} {Process} for {Permissioned} {Blockchain} {Network} ({Hyperledger} {Fabric}),'' in \emph{2017 {IEEE} 36th {Symposium} on {Reliable} {Distributed} {Systems} ({SRDS})}.\hskip 1em plus 0.5em minus 0.4em\relax Hong Kong, Hong Kong: IEEE, Sep. 2017, pp. 253--255. [Online]. Available: \url{http://ieeexplore.ieee.org/document/8069090/}
\BIBentrySTDinterwordspacing

\bibitem{wang_data_2021}
\BIBentryALTinterwordspacing
Z.~Wang, H.~Li, H.~Wang, Z.~Xiao, P.~Lu, Z.~Yang, M.~Zhang, and P.~H.~J. Chong, ``A {Data} {Lightweight} {Scheme} for {Parallel} {Proof} of {Vote} {Consensus},'' in \emph{2021 {IEEE} {International} {Conference} on {Big} {Data} ({Big} {Data})}.\hskip 1em plus 0.5em minus 0.4em\relax Orlando, FL, USA: IEEE, Dec. 2021, pp. 3656--3662. [Online]. Available: \url{https://ieeexplore.ieee.org/document/9671637/}
\BIBentrySTDinterwordspacing

\bibitem{xiao_optimizing_2022}
\BIBentryALTinterwordspacing
Z.~Xiao, H.~Li, H.~Wang, G.~Yang, Q.~Ye, S.~Zou, P.~Lu, and Q.~Lyu, ``Optimizing {Parallel} {Proof} of {Vote} {Consensus} {Based} on {Mimic} {Security} in {Consortium} {Blockchains},'' in \emph{2022 {IEEE} {International} {Conference} on {Big} {Data} ({Big} {Data})}.\hskip 1em plus 0.5em minus 0.4em\relax Osaka, Japan: IEEE, Dec. 2022, pp. 3215--3224. [Online]. Available: \url{https://ieeexplore.ieee.org/document/10021130/}
\BIBentrySTDinterwordspacing

\bibitem{rossi_quantum_2013}
\BIBentryALTinterwordspacing
M.~Rossi, M.~Huber, D.~Bruß, and C.~Macchiavello, ``Quantum hypergraph states,'' \emph{New Journal of Physics}, vol.~15, no.~11, p. 113022, Nov. 2013. [Online]. Available: \url{https://iopscience.iop.org/article/10.1088/1367-2630/15/11/113022}
\BIBentrySTDinterwordspacing

\bibitem{qu_encoding_2013}
R.~Qu, J.~Wang, Z.-s. Li, and Y.-r. Bao, ``Encoding hypergraphs into quantum states,'' \emph{Physical Review A—Atomic, Molecular, and Optical Physics}, vol.~87, no.~2, p. 022311, 2013, publisher: APS.

\bibitem{singh_quantum_2021}
\BIBentryALTinterwordspacing
A.~Singh, K.~Dev, H.~Siljak, H.~D. Joshi, and M.~Magarini, ``Quantum {Internet}—{Applications}, {Functionalities}, {Enabling} {Technologies}, {Challenges}, and {Research} {Directions},'' \emph{IEEE Communications Surveys \& Tutorials}, vol.~23, no.~4, pp. 2218--2247, 2021. [Online]. Available: \url{https://ieeexplore.ieee.org/document/9528843/}
\BIBentrySTDinterwordspacing

\bibitem{wehner_quantum_2018}
\BIBentryALTinterwordspacing
S.~Wehner, D.~Elkouss, and R.~Hanson, ``\BIBforeignlanguage{en}{Quantum internet: {A} vision for the road ahead},'' \emph{\BIBforeignlanguage{en}{Science}}, vol. 362, no. 6412, p. eaam9288, Oct. 2018. [Online]. Available: \url{https://www.science.org/doi/10.1126/science.aam9288}
\BIBentrySTDinterwordspacing

\bibitem{wang_self-tallying_2016}
\BIBentryALTinterwordspacing
Q.~Wang, C.~Yu, F.~Gao, H.~Qi, and Q.~Wen, ``\BIBforeignlanguage{en}{Self-tallying quantum anonymous voting},'' \emph{\BIBforeignlanguage{en}{Physical Review A}}, vol.~94, no.~2, p. 022333, Aug. 2016. [Online]. Available: \url{https://link.aps.org/doi/10.1103/PhysRevA.94.022333}
\BIBentrySTDinterwordspacing

\bibitem{mannalatha_comprehensive_2023}
\BIBentryALTinterwordspacing
V.~Mannalatha, S.~Mishra, and A.~Pathak, ``\BIBforeignlanguage{en}{A comprehensive review of quantum random number generators: concepts, classification and the origin of randomness},'' \emph{\BIBforeignlanguage{en}{Quantum Information Processing}}, vol.~22, no.~12, p. 439, Dec. 2023. [Online]. Available: \url{https://link.springer.com/10.1007/s11128-023-04175-y}
\BIBentrySTDinterwordspacing

\end{thebibliography}

\end{document}